\documentclass[10pt,aps,prb,twocolumn,showpacs,amssymb]{revtex4-1}
\usepackage{amsmath,graphics,epsfig,mathrsfs}
\usepackage{bm}
\usepackage{epigraph}
\usepackage{geometry}                
\usepackage{cases}
\usepackage{cancel}

\usepackage{color}

\usepackage{color}
\definecolor{darkgreen}{rgb}{0.0, 0.5, 0.0}
\definecolor{brown}{rgb}{0.59, 0.29, 0.0}
\definecolor{darkgreen2}{rgb}{0.01, 0.75, 0.24}
\usepackage{hyperref}
\hypersetup{backref=true,pagebackref=true,hyperindex=true,colorlinks=true,citecolor=blue, breaklinks=true,urlcolor=
blue,linkcolor=blue,bookmarks=true,bookmarksopen=true}

\begin{document}
\title{Theory of Topological Spin Hall Effect in Antiferromagnetic Skyrmion: Impact on Current-induced Motion}
\author{C.  A.  Akosa$^{1,2}$}
\email{collins.akosa@riken.jp}
\author{O. A. Tretiakov$^{3}$}
\author{G. Tatara$^1$}
\author{A. Manchon$^2$}
\affiliation{$^1$RIKEN Center for Emergent Matter Science (CEMS), 2-1 Hirosawa, Wako, Saitama 351-0198, Japan}
\affiliation{$^2$King Abdullah University of Science and Technology (KAUST), Physical Science and Engineering (PSE) Division, Thuwal 23955-6900, Saudi Arabia}
\affiliation{$^3$Institute for Materials Research, Tohoku University, Sendai 980-8577, Japan}
\date{\today}

\begin{abstract}   
We demonstrate that the nontrivial magnetic texture of antiferromagnetic skyrmions (AFM-Sks) promotes a non-vanishing topological spin Hall effect (TSHE) on the flowing electrons. This results in a substantial enhancement of the non-adiabatic torque and hence improves the skyrmion mobility.  This non-adiabatic torque increases when decreasing the  skyrmion size, and therefore scaling down results in a much higher torque efficiency. In clean AFM-Sks, we find a significant boost of the TSHE close to van Hove singularity. Interestingly, this effect is enhanced away from the band gap in the presence of non-magnetic interstitial defects. Furthermore, unlike their ferromagnetic counterpart, TSHE in AFM-Sks increases with increase in disorder strength thus opening promising avenues for materials engineering of this effect.     
\end{abstract}
\maketitle

\newpage

As the spintronics community advances the search for high-efficiency, high-density and low power-consuming spintronic devices, alternative materials other than conventional ferromagnets (FMs) are being continuously introduced and explored. Materials with strong spin-orbit coupling have been used extensively to achieve efficient, ultrafast and reliable magnetization switching in magnetic heterostructures via the spin-orbit torques (SOTs) in the presence of an in-plane magnetic field applied collinear to the current \cite{Miron2011, Liu2012b, Garello2014, Liu2012a, Cubukcu2014, Yu2014a}. However, simultaneously applying a magnetic field and current to achieve switching poses technological challenges leading to the quest for alternative material systems.  

Besides FMs, antiferromagnets (AFMs) have recently drawn significant attention \cite{Jungwirth2016, Baltz2017}. The experimental observation of bulk SOTs in locally inversion asymmetric CuMnAs \cite{Wadley2016}, the demonstration of AFM-assisted zero-field SOT switching \cite{Fukami2016, Brink2016}, and the achievement of large anomalous and spin Hall effects in non-collinear AFMs \cite{Surgers2014, Nakatsuji2015, Nayak2016} open promising perspectives for the implementation of AFMs into efficient spin devices. The latter effect is particularly intriguing since it emerges from the coexistence of spin-orbit coupling-driven Berry curvature and non-collinear magnetism. In addition, it has also been predicted that AFM textures such as domain walls driven by SOTs can move much faster than their FM counterparts due to the absence of Walker breakdown \cite{Gomonay2016, Shiino2016}. Therefore, the interplay between topological spin transport and the dynamics of AFM textures is a promising route to explore towards the realization of efficient current-driven control of the AFM order parameter.

Recently, ferromagnetic skyrmions (FM-Sks) have been proposed as good candidates for technological applications due to their weak sensitivity to defects \cite{Lin2013a, Reichhardt2015, Lin2013}, ultra-low critical current density \cite{Lin2013, Iwasaki2013a,Iwasaki2013b,Iwasaki2014,Sampaio2013,Everschor2012,Litzius2017}, enhanced non-adiabatic torque \cite{Akosa2017} and substantial TSHE \cite{Yin2015, Ndiaye2017}. In spite of these remarkable properties, FM-Sks suffer from the so-called skyrmion Hall effect \cite{Litzius2017,Jiang2017,Ado2017}, a motion transverse to the current flow. This parasitic effect hinders the robust manipulation of FM-Sks. In contrast, both analytical theory and micromagnetic simulations recently showed that in AFM-Sks, the skyrmion Hall effect vanishes by symmetry  \cite{Gobel2017, Barker2016,Zhang2016}, which results in an enhanced longitudinal velocity and control efficiency.

In this Letter, we demonstrate that the nontrivial magnetic texture of AFM-Sks promotes a non-vanishing TSHE on the flowing electrons. This results in a substantial enhancement of the non-adiabatic torque and hence improves the skyrmion mobility. This non-adiabatic torque increases when decreasing the skyrmion size, and therefore scaling down results in a much higher torque efficiency.  In clean AFM-Sks, we find a significant enhancement of the TSHE close to van Hove singularity. Most importantly, unlike FM-Sks \cite{Ndiaye2017}, TSHE in AFM-Sks increases in the presence of non-magnetic interstitial defects. Moreover, the TSHE is enhanced away from the band gap in the presence of these defects.


Motivated by the prediction of a metastable single AFM-Sk on a square lattice \cite{Barker2016, Zhang2016,Keesman2016, Fujita2017}, our analysis begins with an isolated  G-type (i.e., checkerboard) AFM-Sk. We consider a system of free electrons coupled via the \textit{s-d} exchange interaction to the magnetic texture ${\bf m}_i$, described by the Hamiltonian 
\begin{equation}\label{eq:tb}
\mathcal{H} = \sum_{i }\varepsilon_i \hat{c}_i^\dagger \hat{c}_i  - t \sum_{\langle ij \rangle} \hat{c}_i^\dagger \hat{c}_j  - J\sum_i \hat{c}_i^\dagger {\bf m}_i\cdot\hat{\boldsymbol{\sigma}}\hat{c}_i, 
\end{equation}
where $J$ is the exchange coupling, $\hat{\boldsymbol{\sigma}}$ is the spin of the electron, $t$ is the nearest-neighbor hopping, $\varepsilon_i$ and $\hat{c}_i^\dagger$  ($\hat{c}_i$) are the onsite energy,  and the spinor creation (annihilation) operator of site $i$ respectively.  For an AFM with equivalent sublattices $a$ and $b$, Eq.~(\ref{eq:tb}) can be rewritten in the coupled sublattice-spin space $\left(| a\rangle, | b\rangle\right) \otimes  \left(| \uparrow\rangle, | \downarrow\rangle\right)$ as
 \begin{equation}\label{eq:gtype}
 \mathcal{H} =  (\varepsilon \hat{\mathbb{I}}_2 + \gamma_k \hat{\tau}_x)\otimes\hat{\mathbb{I}}_2 - J(\hat{\tau}_z\otimes{\bf n} + \hat{\mathbb{I}}_2\otimes{\bf m})\cdot\hat{\boldsymbol{\sigma}}   , 
 \end{equation}
 where $\gamma_k = -2t(\cos k_x a_0 + \cos k_y a_0)$, $a_0$ being the lattice constant, $\hat{\mathbb{I}}_2$ is the $2 \times 2$ identity matrix, ${\bf n} = {\bf m}^a - {\bf m}^b$ and ${\bf m} = {\bf m}^a + {\bf m}^b$ are unit vectors in the direction of the N\'eel vector, and the total magnetization, respectively.  $\hat{\boldsymbol{\tau}}$ and $\hat{\boldsymbol{\sigma}}$ are the Pauli matrices of the sublattice  and spin subspaces, respectively. The eigenvalues and eigenstates corresponding to Eq.~(\ref{eq:gtype}), in the limit of smooth and slow varying magnetic texture, are given by \cite{Saidaoui2017}
 \begin{subequations}
\begin{eqnarray}\label{eq:disp}
\varepsilon_s^\eta(k) &=& s\sqrt{\gamma_k^2 + J^2}, \\
\Psi_s^{\eta, \sigma} &=& \sum_\eta s \sqrt{\frac{1 + s\sigma P_{k}^\eta}{2}} |\eta\rangle\otimes|\sigma \rangle,
\end{eqnarray}\label{eq:wf}
\end{subequations}
where $s =+1(-1)$ represents the bands above (below) the band gap and $ P_{k}^\eta = \eta J/\sqrt{\gamma_k^2 + J^2}$ is the polarization of the local density of states of the $\eta$-sublattice.  Without loss of generality, we place the Fermi energy in the bottom band, i.e. we set $s = -1$ in the rest of the paper.  A unitary transformation that acts \textit{only} on the spin-subspace $\mathcal{U} = e^{-i\frac{\theta}{2} \boldsymbol{\sigma}\cdot {\bf e}_\phi}$, where ${\bf e}_\phi = {\bf z}\times{\bf n}/|{\bf z}\times{\bf n}|$ \cite{Akosa2017, Barnes2007,Tserkovnyak2008, Zhang2009, Bisig2016}, transforms Eq.~(\ref{eq:gtype}) into
\begin{equation}\label{eq:ha}
\tilde{\mathcal{H}}  \approx \frac{(\hat{\bf p} - e\hat{\bf A})^2}{2m}\hat{\tau}_x\otimes\hat{\mathbb{I}}_2  -  J\hat{\tau}_z \otimes\hat{\sigma}_z  +  e\hat{\mathbb{I}}_2\otimes\hat{V}.
\end{equation}
$\hat{\bf A} = -\frac{\hbar}{2e}[\hat{\boldsymbol{\sigma}}^\eta\cdot ({\bf n}\times\partial_i{\bf n})]{\bf e}_i$ and $\hat{V}= \frac{\hbar}{2e}[\hat{\boldsymbol{\sigma}}^\eta\cdot ({\bf n}\times\partial_t{\bf n})]{\bf z}$ are the vector and scalar potentials respectively. $\hat{\boldsymbol{\sigma}}^\eta = \frac{1}{2}(\hat{\mathbb{I}}_2  +\eta \hat{\tau}_z)\otimes\hat{\boldsymbol{\sigma}}$ is the spinor operator, with $\eta = +(-)1$ for the $(a)b$-sublattice.  As a result, the spin-dependent carriers feel an emerging electromagnetic field for the $\eta$-sublattice  given by,
\begin{subequations}\label{eq:em_fd}
\begin{equation}
{\bf E}^{\eta, \sigma}_{\rm em} = (\sigma\hbar/2e)\mathcal{P}^\eta_\sigma [(\partial_t{\bf n}\times\partial_i{\bf n})\cdot{\bf n}] {\bf e}_i , 
\end{equation}
\begin{equation}
{\bf B}^{\eta, \sigma}_{\rm em} = -(\sigma\hbar/2e)\mathcal{P}^\eta_\sigma [(\partial_x{\bf n}\times\partial_y{\bf n})\cdot{\bf n}]{\bf z},
\end{equation}\label{eq:emh}
\end{subequations}
where $\mathcal{P}^\eta_\sigma = 1 - \sigma P_{k}^\eta$, $\sigma = +(-)1$ for $\uparrow(\downarrow)$ spin. 

Insight into the similarities and differences in the physics of emergent electrodynamics in FM-Sks and AFM-Sks can be inferred from the pre-factor $\mathcal{P}^\eta_\sigma$. 
Indeed, unlike FM-Sks in which electrons feel an emergent electromagnetic field of \textit{opposite sign} for different spins, 
the \textit{magnitude} of these emergent fields in AFM-Sks is both spin- and sublattice-dependent, as a result, strongly depends on the dispersion. Under the action of an external electric field {\bf E} along {\bf x}, i.e. ${\bf E} = E{\bf x}$, the resulting spin- and sublattice-dependent local carrier current density is given by 
\begin{equation}\label{eq:lscc}
{\bf j}^{\eta, \sigma}_{\rm cc} = \sigma_{0}^{\eta, \sigma}{\bf E} + \sigma_{0}^{\eta, \sigma}{\bf E}^{\eta, \sigma}_{\rm em}+\frac{\sigma_{\rm H}^{\eta, \sigma}}{B_{\rm em}^{\eta,\sigma}}{\bf E} \times{\bf B}^{\eta, \sigma}_{\rm em},
\end{equation}
where $B_{\rm em}^{\eta, \sigma}$ is a constant with the dimension of magnetic induction, $\sigma_{0}^{\eta, \sigma}$  and $\sigma_{\rm H}^{\eta, \sigma}$ are spin-$\sigma$ contribution of the normal and Hall conductivity, respectively. The local charge and spin current densities are calculated as \cite{Akosa2017}
\begin{subequations}\label{eq:top_curr}
\begin{eqnarray} \label{eq:top_je} \nonumber
{\bf J}^\eta_{e} &=&  \sigma_{0}^\eta  E \left({\bf x}  -  \eta (1 - P_{k}^\eta P_H^\eta){\lambda_{\rm H}^\eta}^2 \mathcal{N}_{x,y} {\bf y}  \right) \\
&& + (\hbar/2e) \left(P_0^\eta -  P_{k}^\eta \right)\sigma_{0}^\eta \mathcal{N}_{t, i} {\bf e}_i ,
\end{eqnarray}
\begin{eqnarray}\label{eq:top_js}\nonumber
{\bf J}^{\eta}_{s} &=& b_{\rm J}^\eta  M_s {\bf n}\otimes\left( {\bf x} -  \eta P_t {\lambda_{\rm H}^\eta}^2 \mathcal{N}_{x,y}{\bf y} \right)   \\ 
&&   +  P_d {\lambda_{\rm E}^\eta}^2 \mathcal{N}_{t,i} M_s{\bf n}\otimes{\bf e}_i ,
 \end{eqnarray}
\end{subequations}
where $P_t =  (P_H^\eta - P_{k}^\eta)/P_0^\eta$ , $P_d =  (1 - P_{k}^\eta P_0^\eta)$, $\mathcal{N}_{\nu,\mu} = {\bf n}\cdot(\partial_\nu{\bf n}\times\partial_\mu{\bf n})$, with $\nu,\mu \in(x,y,t)$, $b_{\rm J}^\eta = \hbar P_0^\eta \sigma_0^\eta E/ 2m_eM_s$, $\lambda_{\rm H}^\eta = \sqrt{\hbar \sigma_{\rm H}^{\eta}/2e\sigma_{0}^{\eta} B_{\rm em}}$  and $\lambda_{\rm E}^{\eta} = \sqrt{\hbar^2\sigma_{0}^{\eta}/4e^2M_s}$  are the length scales associated with the emergent magnetic and electric fields respectively \cite{Akosa2017}.
\begin{figure}[t!]
\includegraphics[width=7.8cm]{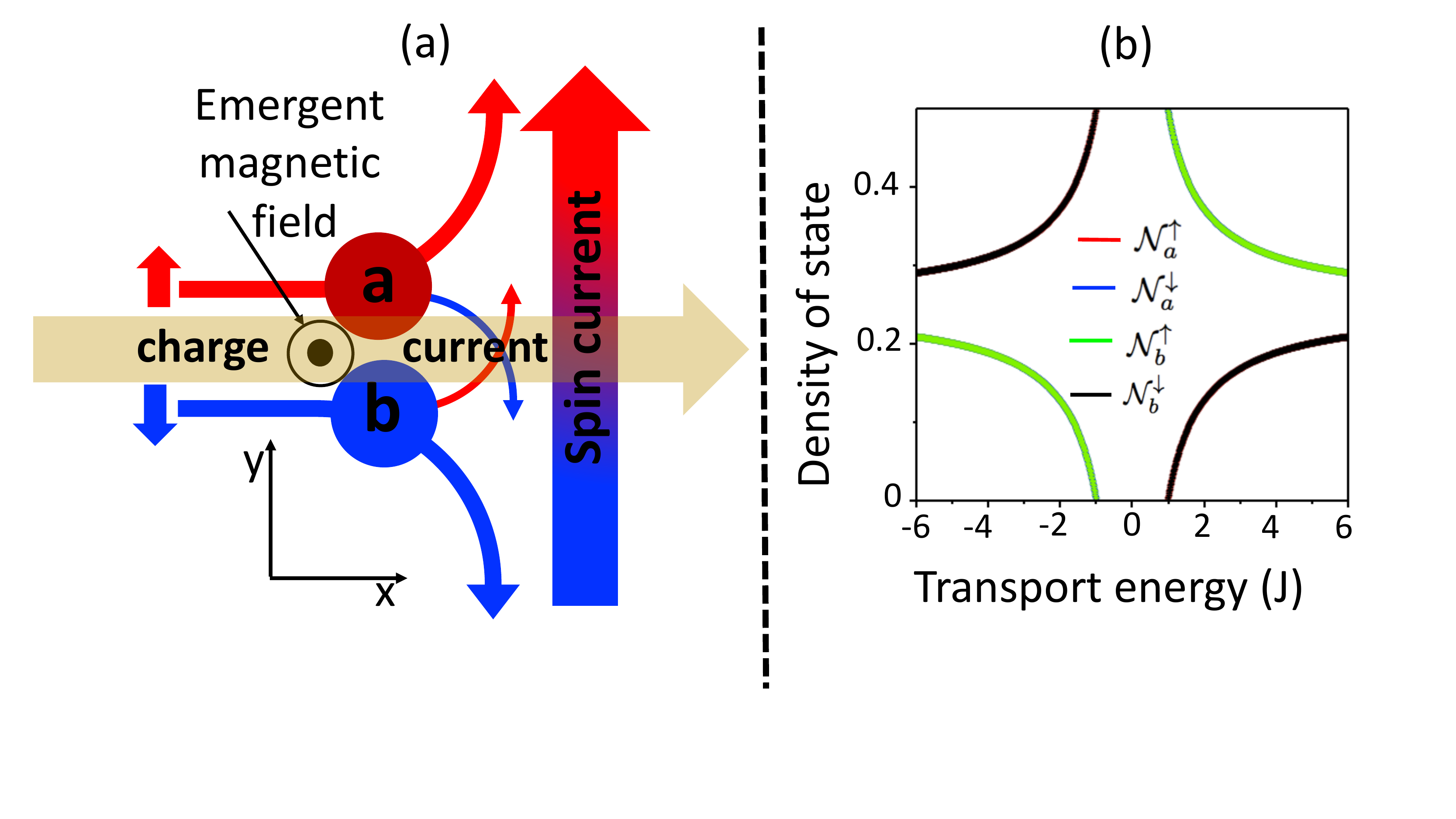}
\caption{(a) Schematic diagram showing the physical origin of TSHE due to the emergent magnetic field in AFM-Sk. (b) Enhancement of TSHE close to the band gap due to the relative population of electrons on the sublattices.}\label{fig:spincurrent}
\end{figure}

Interesting physics of charge and spin transport in AFM-Sks can be deduced from Eq.~(\ref{eq:top_curr}). Indeed, since $\eta$,  $b_{\rm J}^\eta$, $P_{k}^\eta$ and $P_{\rm 0(H)}^\eta $ change their sign on different sublattice, (i) there is no macroscopic  transverse (\textit{along {\bf y}}) charge current, i.e. no topological Hall effect (THE) \cite{Barker2016, Zhang2016}, (ii) even though there is no macroscopic longitudinal (\textit{along {\bf x}}) \textit{spin} current, there is a finite transverse spin current i.e. finite TSHE \cite{Buhl2017}.  The TSHE is proportional to the polarization of the local density of state $P_k$ and since $P_k \ne 0$ and increases reaching a maximum at the band gap, the TSHE is expected to increase accordingly.  The physical origin of this TSHE stems from an interplay between the emergent magnetic field and the dispersion of the underlying system. Indeed, since the emergent magnetic field given by Eq.~(\ref{eq:emh}) deflects electrons with opposite spins in opposite directions, coupled to the two-fold degeneracy inherent in AFMs with equivalent sublattices, a continuous transverse \textit{pure} spin current flows in the system as depicted in Fig.~\ref{fig:spincurrent} \textcolor{blue}{(a)}. 

\begin{figure}[t!]
\includegraphics[width=7.2cm]{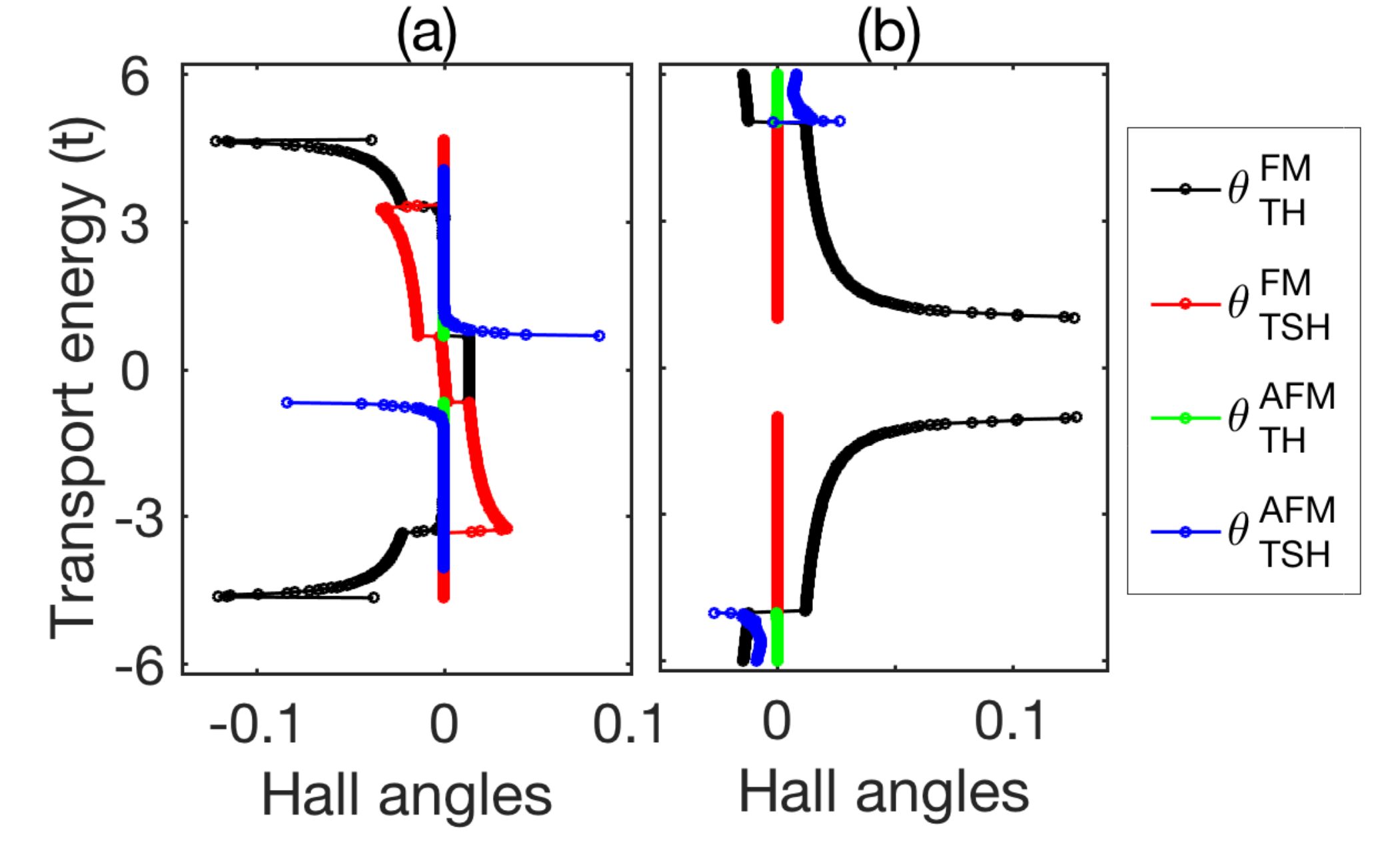}
\caption{Computed THE and TSHE for FM-Sk and AFM-Sk (quantified by their respective angles, $\theta_{\rm TH}$ and $\theta_{\rm TSH}$) as a function of the Fermi energy in the intermediate (a) with $J = 2t/3$ and strong (b) with $J = 5t$, exchange limits. The skyrmion radius is set to $r_0 = 8a_0$.}\label{fig:ballis}
\end{figure}

Our theoretical predictions are verified by means of a tight-binding model of an isolated AFM-Sk on a square lattice of size $82\times82   a_0^2$  (i.e. $41\times41$ AFM unit cells), described by Eq.~(\ref{eq:tb}). Using the KWANT code \cite{Groth2014}, we investigate the topological Hall transport of this system by means of a four-terminal system\cite{Ndiaye2017} in both the intermediate ($J = 2t/3$) and strong ($J = 5t$) exchange limits \cite{Yin2015} and compare with an equivalent FM-Sk.

In a clean system in both the intermediate and strong antiferromagnetic limits as shown in Fig.~\ref{fig:ballis} \textcolor{blue}{(a)} and \textcolor{blue}{(b)}, we find finite TSHE with a substantial increase close to the band gap in AFM-Sk [blue line in Fig.~\ref{fig:ballis} \textcolor{blue}{(a)} and \textcolor{blue}{(b)}], which can be much larger than in FM-Sk. And since this effect increases with the skyrmion density \cite{Ndiaye2017}, this can result in a substantial TSHE capable of inducing magnetization dynamics and/or switching on an adjacent attached FM layer. Furthermore, unlike FM-Sks, AFM-Sks exhibit no THE due to the cancellation of the charge current contributions from both sublattices \cite{Barker2016} [green line in Fig.~\ref{fig:ballis} \textcolor{blue}{(a)} and \textcolor{blue}{(b)}].  Our numerical results are consistent with our analytical predictions Eq.~({\ref{eq:top_curr}).

\begin{figure}[t!]
\includegraphics[width=7.5cm]{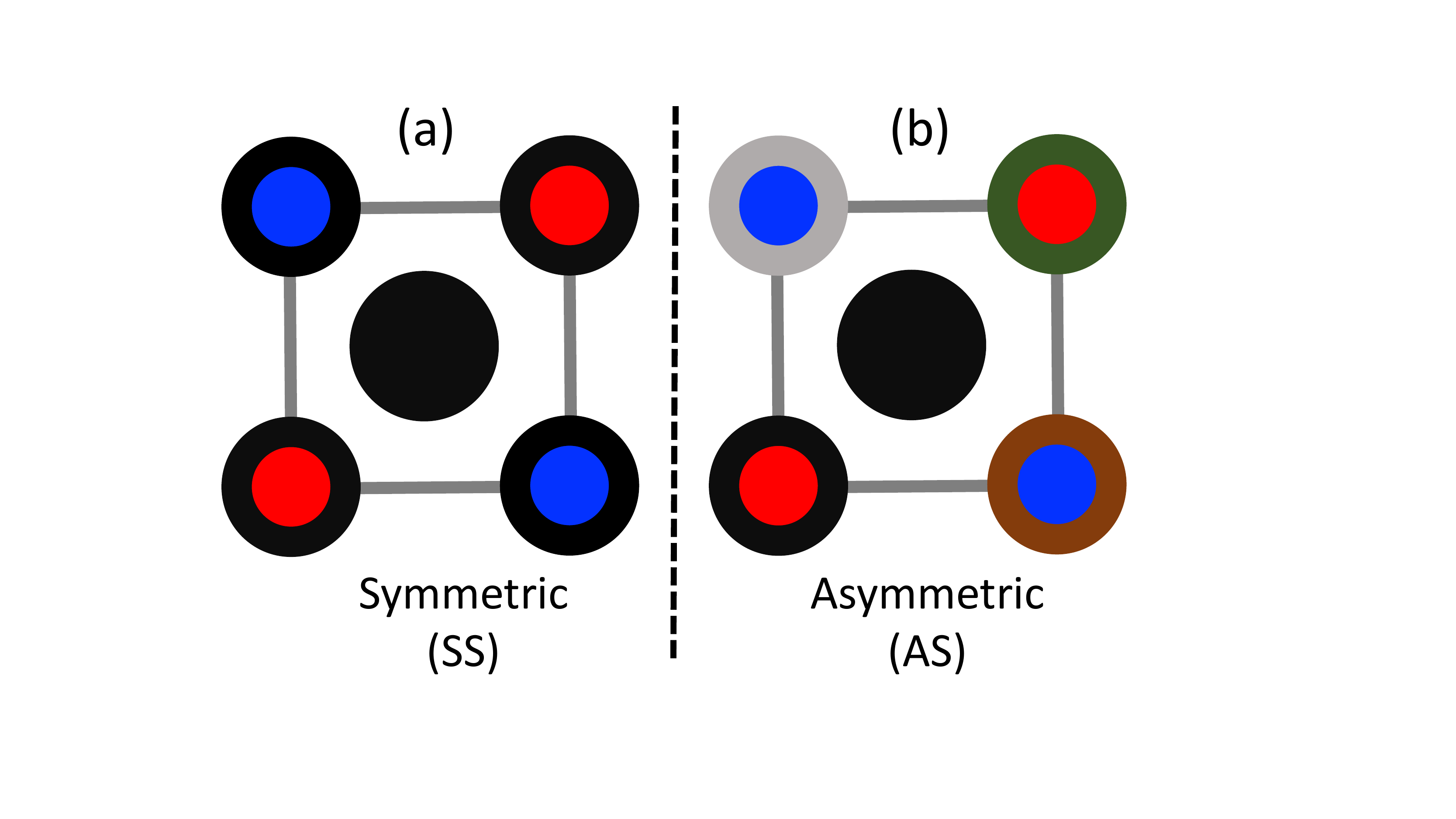}
\caption{Illustration of different effects of an interstitial defect (black) in an antiferromagnet with equivalent sublattices \textit{a} (red) and \textit{b} (blue). 
The color dressings represent relative change of the onsite energy due to the defect. This dressing can be (a) symmetric within the unit cell, or (b) asymmetric.}\label{fig:disorder}
\end{figure}

Next, we investigate the impact of impurities which are omnipresent in real materials. We focus on non-magnetic \textit{interstitial} defects, which can have a symmetric (SS) or an asymmetric (AS) effect, as illustrated in Fig~\ref{fig:disorder}. The defects are modeled as a random onsite energy $\varepsilon_i = \mathcal{V}_i$, where $\mathcal{V}_i\in \left(-\frac{_W}{^2}, \frac{_W}{^2}\right)$,  $W$ defines the strength of the disorder assigned to a site (AS) or unit cell (SS), and to ensure convergence, we average over 10000 disorder configurations. As shown in Figs.~\ref{fig:compare} \textcolor{blue}{(a)} and \textcolor{blue}{(b)}, for FM-Sk the presence of disorder progressively quenches the TSHE (black curve); in contrast, in the case of AFM-Sk only AS disorder, i.e. disorder that induces decoherence within the antiferromagnetic unit cell -- quenches the TSHE (red curve). In fact, SS disorder, that preserves coherence between the two sublattices, substantially enhances the TSHE for moderate disorder (blue curve).

To conclude this study, we investigate the impact of the topological spin current in Eq.~(\ref{eq:top_js}) on the dynamics of an AFM-Sk. To achieve this, we calculate the corresponding spin torque as ${\bf T}_t^\eta = -\boldsymbol{\nabla}\cdot{\bf J}_{s}^\eta$ and for the sake of completeness, we include non-adiabatic effects arising from, for example spin relaxation \cite{Zhang2004} via a constant nonadiabaticity parameter $\beta$. This yields a total spin transfer torque ${\bf T}^\eta = -{\bf T}_t^\eta - \eta \beta M_s b_J^\eta{\bf n}\times\partial_x{\bf n}$ given as
\begin{eqnarray}\label{eq:top_tot} \nonumber
{\bf T}^\eta &=&  M_sb_{\rm J}^\eta \left[ \partial_x {\bf n}  -  \eta \beta {\bf n}\times\partial_x{\bf n}\right] \\ 
&&  M_s {\bf n}\times\left[ \alpha^\eta_{\rm T}({\bf r}) \partial_t{\bf n} -  \beta^\eta_{\rm T}({\bf r})b_J^\eta \partial_x{\bf n}\right],
\end{eqnarray}
where $\alpha^\eta_{\rm T}({\bf r}) $  and $\beta^\eta_{\rm T}({\bf r}) $ are proportional to the topological contribution to the damping  and non-adiabatic torques given by
\begin{subequations}
\begin{eqnarray}
\alpha^\eta_{\rm T}({\bf r}) &=& pq P_d{\lambda_{\rm E}^\eta}^2 \mathcal{N}_{x,y}, \\
\beta^\eta_{\rm T}({\bf r }) &=& \eta pq P_t {\lambda_{\rm H}^\eta}^2  \mathcal{N}_{x,y}.
\end{eqnarray}
\end{subequations}

\begin{figure}[t]
\includegraphics[width=7.8cm]{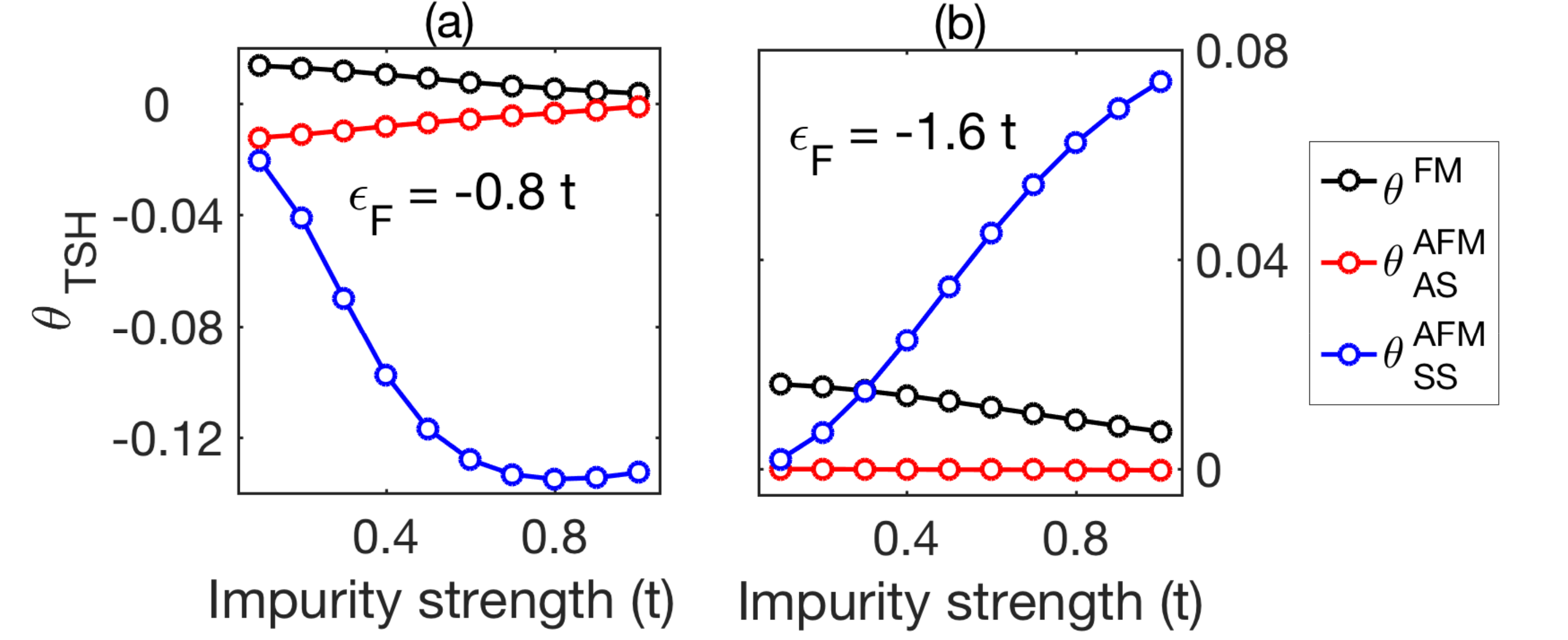}
\caption{Dependence of TSHE on the strength of disorder for both FM-Sk and AFM-Sk: (a) close to the band gap and (b) around the middle of the band. Notice the substantial enhancement of TSHE in the presence of impurity in the case of symmetric (SS -- blue curve) compared to asymmetric (AS -- red curve) defects in the intermediate exchange limit, i.e. $J = 2t/3$. }\label{fig:compare}
\end{figure}

From Eq.~(\ref{eq:top_tot}), it appears clear that just as in FM-Sks \cite{Bisig2016, Akosa2017}, the transverse topological spin current flowing in AFM-Sks directly enhances the non-adiabatic torque and the damping. 
Moreover, this non-adiabatic torque being topological in nature, it increases when decreasing the skyrmion size. As a result, the current-driven efficiency increases when the skyrmion becomes smaller. 
Following the standard theoretical scheme to study the dynamics of antiferromagnetic textures \cite{Swaving2011, Hals2011, Swaving2012, Tveten2013, Rodrigues2017} supplemented by the topological torques in Eq.~(\ref{eq:top_tot}), we obtain the equation of motion of the N\'eel order parameter as 
\begin{equation}\label{eq:llg_fin}
\frac{1}{\bar{a}\tilde{\gamma}}\partial_t^2{\bf n}   + \alpha_{\rm eff}({\bf r}) \partial_t{\bf n}     =   \gamma{\bf f}_{\bf n}  +  \beta_{\rm eff}({\bf r}) b_{\rm J}\partial_x{\bf n},
\end{equation}
where $\tilde{\gamma} = \gamma/(1+\alpha^2)$, $\alpha_{\rm eff}({\bf r})  = \alpha + \alpha_{\rm T}({\bf r}) $ and $\beta_{\rm eff}({\bf r})  = \beta + \beta_{\rm T}({\bf r}) $.  
We consider an AFM-Sk with profile defined in spherical coordinates as ${\bf n} = (\cos\Phi\sin\theta, \sin\Phi\sin\theta, \cos\theta)$, where $\cos\theta = p(r_0^2 - r^2)/(r_0^2 + r^2)$ and $\Phi = qArg(x+iy) + c\pi/2$, with $p$, $q$, and $c$ being the polarization, vorticity, and chirality of the texture, respectively, which take values $\pm 1$ \cite{Akosa2017}. Using this profile, the terminal velocity of the structure is given as
\begin{equation}
 v_y = 0   \hspace{4 mm} \mbox{and} \hspace{4 mm}  v_x = (\beta_{\rm eff}/\alpha_{\rm eff}) b_{\rm J},
 \end{equation}
where $\beta_{\rm eff} = \beta + \frac{4P_t \mathcal{S}_2}{3}\frac{\lambda_{\rm H}^2}{r_0^2}$ and $ \alpha_{\rm eff} = \alpha + \frac{4P_d\mathcal{S}_2}{3}\frac{\lambda_{\rm E}^2}{r_0^2}$, 
$r_0$ being the radius of the skyrmion. $\mathcal{S}_2 = \sum_{k = 0}^2 (r_0^2/(r_0^2 + R^2))^k$ is a geometric factor which equals unity in the limit $R\gg r_0$. To provide a qualitative estimate, using realistic materials parameters $M_s = 800 \mbox{   KA/m}$, $\alpha = 0.01$, $\beta = 0.02$, $\sigma_H/\sigma_0 = 0.1$, $\sigma_0 = (\mbox{4  $ \mu \Omega$  cm})^{-1}$, $P_0 = 0.7$, $P_k = 0.35$, $B_{\rm em} = \mbox{2.5 T}$, and $J_e = 5\times 10^{11}$ \mbox{A/m$^2$}, we obtain $\lambda_E^2$ = 0.4\mbox{ nm}$^2$ and $\lambda_H^2$ = 13.2\mbox{ nm}$^2$, which translates to a velocity of up to $393 \mbox{ m/s}$ for skyrmion size of $10 \mbox { nm}$.

Micromagnetic simulations originally predicted that skyrmions have in principle limited sensitivity to local and edge defects owing to very weak interactions \cite{Lin2013a, Reichhardt2015} and their finite spatial extension \cite{Iwasaki2013b, Sampaio2013}. Indeed, the ability of a defect to pin a skyrmion increases when the size of the skyrmion becomes comparable to the size of the defect \cite{Lin2013a}. Hence, scaling down the skyrmion towards sub-100 nm size results in low skyrmion mobility and large critical depinning currents in polycrystalline systems \cite{Legrand2017}. What makes AFM-Sks remarkable in this respect is the fact that the torque efficiency itself increases when reducing the skyrmion size, as discussed above. While this topological torque only contributes to the transverse motion of FM-Sks, it drives the longitudinal motion of AFM-Sks and therefore directly competes with the enhanced pinning potential. This unique property could be a substantial advantage to compensate the increasing pinning upon size reduction. Furthermore, our calculations show that TSHE is enhanced in the presence of moderate disorder that is omnipresent in real materials, demonstrating the robustness of the proposed approach for device applications.

This work was supported by Grant-in-Aid for Scientific Research (B) No. 17H02929, from the Japan Society 
for the Promotion of Science and Grant-in-Aid for Scientific Research on Innovative Areas No. 26103006 from The Ministry of Education, Culture, Sports, Science and Technology
(MEXT). C. A. A.  and A. M. acknowledges support from King Abdullah University of Science and Technology (KAUST). O.\,A.\,T. acknowledges support by the Grants-in-Aid for Scientific Research (No.\,25247056, No.\,17K05511, and No.\,17H05173) from the MEXT of Japan, MaHoJeRo (DAAD Spintronics network, Project No. 57334897), and by JSPS and RFBR under the Japan-Russia Research Cooperative Program.


\end{document}